\documentclass[twocolumn,showpacs,aps]{revtex4}
\usepackage[dvips]{graphicx}
\usepackage{bm}

\begin{document}
\date{\today}

\title{Surface Instability of Icicles}
\author{Naohisa Ogawa and Yoshinori Furukawa
\footnote{ogawa@particle.sci.hokudai.ac.jp,~frkw@lowtem.hokudai.ac.jp}}
\affiliation{Institute of Low Temperature Sciences, \\
Hokkaido University, Sapporo 060-0819 Japan}
\begin{abstract}
Quantitatively-unexplained stationary waves or ridges often encircle icicles.
Such waves form when roughly 0.1 mm-thick layers of water flow down the icicle.
These waves typically have a wavelength of 1cm approximately independent of external temperature, icicle thickness, 
and the volumetric rate of water flow.
In this paper we show that these waves can not be obtained by naive Mullins-Sekerka instability, 
but are caused by a quite new surface instability related to the thermal diffusion and hydrodynamic effect 
of thin water flow.
\end{abstract}
\maketitle

\section{Introduction}
Interesting wave patterns often form on growing icicles that are covered with a 
thin layer of flowing water (See Fig.1)\cite{icicle}. 
For many of these patterns, the wavelength has a Gaussian distribution centered at 
approximately 8 mm; however, despite their common occurrence, 
there is no quantitative explanation for this wavelength distribution \cite{icicle},\cite{matsu}. 
These waves are associated with the growth of the icicles and the flow of fluid along the icicle. 
Hence, there are several processes occurring that should be considered. 
These include crystallization from the melt, latent heating at the ice-melt interface, laminar flow with two interfaces 
(ice-melt and melt-air), evaporation of liquid, and transport of heat through the surrounding air. 
That the waves tend to encircle the icicle clearly indicates the importance of gravity-induced flow, 
although the specific interactions between the flow, 
the ice growth, and the heat flow through both interfaces must be considered. 

\begin{figure}[h]
\begin{center}
\includegraphics[width=5cm]{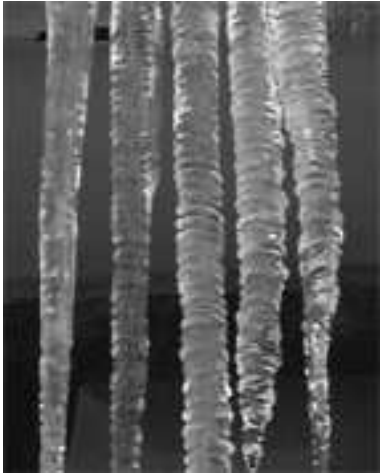}
\end{center}
\caption{Waves on icicle.}
\end{figure}

  In the study of crystal growth, such a surface instability is usually explained by Mullins-Sekerka theory
(hereafter abbreviated as MS). 
MS theory is based on two observations: Laplace instability and Gibbs-Thomson effect.
(For detailed explanation,  refer to text books and original paper by Mullins and Sekerka cited in \cite{MS}.)

 To a good approximation, the ice in the icicle has a uniform temperature of 273 K, 
 thus temperature gradients into the ice are insignificant, 
 and the external temperature field is time independent and satisfies Laplace's equation. 
We further assume that the external temperature is below 273 K; i.e., the ice is not melting on average. 
At a convex point, the temperature gradient is higher.
Because heat flow is proportional to the gradient of temperature, 
the larger heat flow at a convex point rapidly removes
 latent heat from the ice surface thus allowing the convex points to increase in size rapidly. 
Conversely, concave regions grow relatively slowly.
This phenomenon suggests that short-wavelength fluctuations increase in amplitudes more rapidly 
than long wavelength fluctuations.
We refer to this as the Laplace instability.

Next we explain the Gibbs-Thomson effect. 
The surface of a solid object has its own energy per area called surface free energy. 
If a molecule attaches itself to the surface near a convex point, the surface area increases, 
resulting in an increase in energy. 
On the other hand, if a molecule becomes attached to the surface near a concave point, 
absorption of the molecule makes the surface area smaller. 
Therefore, absorption of molecule at a concave point is more energy-efficient 
than is absorption at a convex point. For this reason, the melting point depends on the curvature of an object, i.e., 
the shape of the surface area. 
The melting point is lower at a convex surface (easy to melt) and is higher at a concave surface 
(hard to melt). Such an effect suppresses the fluctuation and makes surface flat. 
This is called Gibbs-Thomson effect, which is opposite to that of Laplace instability.
Laplace instability enhances shorter wavelength fluctuation, 
and Gibbs-Thomson effect suppresses shorter wavelength fluctuation. 
From these two effects, we have fluctuation of specific wavelength mainly. 
These two effects are incorporated in Mullins-Sekerka's theory \cite{MS},\cite{FK}. 
However, the simple application of this theory cannot take place in the case of icicles.
The reasons are as follows.:
Firstly, the water layers on icicles are too thin to produce a Laplace instability.
Because the instability requires that the water thickness be larger than the wavelength of the fluctuation.
Secondly, the curvature is too small. (Wavelength 1 cm is much larger than 10 $\mu m$ 
which GT effect requires to be effective.)
Therefore the nature of wave patterns along icicles can not be explained by naive MS theory. \\
 
We neglect the fluid instabilities that require turbulence 
because there should be no turbulence in these water films \cite{Fluid}.
This is because the layers are only about 100 $\mu m$ thick and the flow speeds are about 2-4cm/s;
the resulting Reynolds numbers are only about 1 and the flow is laminar.
The hydrodynamics in thin water layer is also discussed by Wettlaufer et.al 
to explain the premelting dynamics, but the discussion here is essentially different \cite{Wettlaufer}.

For our analysis, we assume water flow on a ramp (See Fig.2) because it is
simpler to treat and relevant experiments for this geometry are available \cite{matsu}.
Much of the same processes and relative length scales occur in both system because
 the water-layer thickness ($\sim 10^{-4}m$) is much smaller than the radius of icicle ($\sim 10^{-2}m$).
Furthermore, Matsuda \cite{matsu}, observed  wave patterns on such an ice ramp; 
for example, at $\theta = \pi/2$, the wavelength was about 8 mm.

\begin{figure}[h]
\begin{center}
\includegraphics[width=6cm]{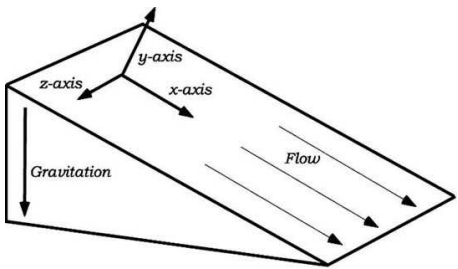}
\end{center}
\caption{Flow on a ramp inclined at $\theta$ degrees.}
\end{figure}

Liquid flow of a thin water on a flat ramp(Benney's liquid film) produces waves
 \cite{Benney}(see also \cite{Landau}); however, these waves travel down the ramp and thus are unlike
the case on icicles. Due to the explicit calculation, the wavelength of these travelling surface waves is about 1 cm,
which agrees with wavelength along icicles, but they move at about 4-8 cm/s which is twice the speed of the fluid.
Benney's wave is caused by gravity and surface tension, but it is unclear how it applies to the standing waves on ice 
unless the travelling waves can become pinned to a fixed location; 
such a pinning mechanism has not yet been proposed.

Our  approach is to assume static flow with small ripples on the ramp surface, 
and then calculate the growth rate for the ripples by solving the thermal diffusion equation in the background fluid.
In section 2, we discuss the fluid dynamics of a thin water flowing along a ramp, and then
in section 3, we couple the thermal diffusion process to the flow.
The thermal diffusion in air is solved in section 4, and in section 5 we discuss the growth rate of fluctuation on icicles 
by combining the solutions for thermal diffusion equations in two regions; air and water.

\section{Hydrodynamics in a thin-layer of water}

   We consider the fluid mechanics of a thin water layer with depth $h(x)$ as sketched in Fig.3.
Over each wavelength, the average depth is $h_0$.
There are two material boundaries, one is the solid-liquid boundary SL, and the other is the air-liquid boundary AL.
The x-axis is along the ramp and increases in the downhill direction, whereas  y-  is the outward normal to the ramp.
The SL surface is not flat and given by

\begin{equation}
y=\phi(x), ~~~~ \phi(x) = \delta \sin kx.
\end{equation}

AL surface is given by
\begin{equation}
y= \xi(x) \equiv \phi(x) + h(x),~ \mbox{ with}~ <h(x)> = h_0,
\end{equation}
where $<h>$ means average over wavelength in x-direction.

\begin{figure}[t]
\begin{center}
\includegraphics[width=5cm]{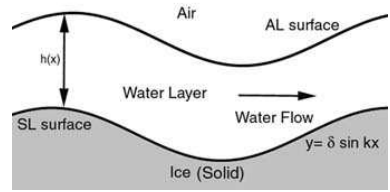}
\end{center}
\caption{Two boundaries: SL and AL.}
\end{figure}

In the experimental data \cite{matsu}, surface velocity of fluid is about 3cm/s, 
and by using  Nusselt equation which is shown later \cite{Landau}, 
the average water-layer thickness $h_0$ was around $0.1$ mm.
The wavelength  $\lambda = 2\pi/k$ is about 1 cm experimentally.
We use the parameter
\begin{equation}
\mu \equiv kh_0 ,
\end{equation}
which is $ 6 \times 10^{-2}$ for the typical experimental values above. 
In general, having $\mu << 1$ defines the long-wavelength approximation.

Our key assumptions are as follows.

\begin{itemize}
\item Steady-state(time-independent) flow.
\item  $\mu <<1$ (long wavelength approximation)
\item Incompressible  fluid.
\end{itemize}

We prefer to use the following dimension-less variables.
For length,
\begin{eqnarray}
x &\to& x_* \equiv kx,\\
y &\to& y_* \equiv y/h_0.
\end{eqnarray}
The thickness of water layer is thus
\begin{equation}
h(x) \to h^*(x)=h(x)/h_0 = 1 + \tilde{h}(x).
\end{equation}
And the height of  the SL and AL surfaces are

\begin{eqnarray}
SL: ~~\phi_*(x) &=& (\delta/h_0) \sin x_* \equiv \eta \sin x_*,\\
AL: ~~\xi_*(x) &=& h_*(x) + \phi_*(x)= 1+\tilde{h}(x)+\eta \sin x_*.
\end{eqnarray}
The characteristic flow velocity  in the x-direction is 

\begin{equation}
U_0 = \frac{gh_0^2 \sin \theta}{2 \nu},
\end{equation}
where $g$ is the gravitational acceleration and the viscosity $\nu = 1.8 \times 10^{-6} m^2/s$ at 
0 degrees centigrade.
This is a Nusselt equation, the theoretically predicted velocity at AL surface when $\delta=0$, i.e., 
for a flat SL surface with uniform thickness \cite{Landau}.
This unperturbative solution  results by equating the gravitational force to the viscous force.
The velocity distribution is parabolic in y;
\begin{equation}
v_x = U_0 (2\frac{y}{h_0} - (\frac{y}{h_0})^2).
\end{equation}
We consider perturbations of this solution.
By using the above formula for the speed, we  relate the experimentally-determined flow $Q$ to $h_0$ and, 
surface velocity $U_0$.
The flow quantity is defined by $Q =2\pi R \bar{U} h_0$,
where $\bar{U}= \frac{1}{h_0} \int_0^{h_0} v_x dy = 2U_0/3$, the mean speed; R, the radius of icicle.
In the experiment \cite{matsu}, Matsuda used the flow $ Q =160 ml/hr$
and width of gutter $l= 3cm$ ($l$ corresponds to $2\pi R$) because this produced the clearest  waves.
This gives $\bar{U} h_0 = 1.48 \times 10^{-6} m^2/s$. 
From $\bar{U}= 2U_0/3$ and $U_0 = \frac{gh_0^2}{2\nu}$ (by setting $\theta = \pi/2$),
we get $U_0 =2.4 \times 10^{-2} m/s$ with $h_0 = 0.93 \times 10^{-4} m$.
On the other hand, his measurement of the surface mean velocity by observing the motion of  fine grain was
   $U_0 =4 \times 10^{-2} m/s$ at $\theta = \pi/2$.
Hence, we assume $U_0 = (2.4 \sim 4) \times 10^{-2} m/s$
with $h_0 = (0.93 \sim 1.21) \times 10^{-4} m$
as the experimental surface speed and water-layer thickness.

In the y-direction, characteristic velocity is
\begin{equation}
V_0 = \mu U_0.
\end{equation}
We write $u$ as speed in the x-direction, $v$ as that in the y-direction,
and $P$ as the pressure.
Dimensionless speeds and pressure are given by
\begin{eqnarray}
u_* &=& u/U_0,\\
v_* &=& v/V_0,\\
P_* &=& \frac{P}{\rho g h_0 \sin \theta}.
\end{eqnarray}
Other dimensionless constant parameters are, respectively, the Reynolds number and the Weber number
\begin{eqnarray}
R &\equiv& \frac{h_0 U_0}{\nu},\\
W &\equiv& \frac{\gamma}{\rho g h_0^2}.
\end{eqnarray}
where $\gamma \sim 7.6 \times 10^{-2}N/m$ is the surface tension of liquid water.
Approximate values of these quantities, $U_0 \sim 3 \times 10^{-2}m/s$, $h_0 \sim 10^{-4}m$, and
$\nu \sim 1.8 \times 10^{-6}m^2/s$ predict $R \sim 1.5$ and $W \sim 10^3$.
This value of Reynolds number indicates laminar flow.
The flow components u and v satisfy the steady-state  Navier Stokes equation for incompressible fluids:
\begin{equation}
  ({\bf v} \cdot \nabla)  {\bf v}  =  -\frac{\nabla P}{\rho } + {\bf g} +
  \nu \Delta {\bf v}. \label{eq:NS}
\end{equation}

The incompressibility condition is
\begin{equation}
\nabla \cdot {\bf v} = 0.
\end{equation}
Mass conservation law at the water-air interface requires
\begin{equation}
\frac{d \xi}{d x}  = \frac{v(x,\xi(x))}{u(x,\xi(x))}.
\end{equation}
The boundary conditions at the SL surface are zero fluid velocity:
\begin{equation}
u(x, y= \phi(x)) = 0, ~~~~v(x, y= \phi(x)) = 0.
\end{equation}

Stress balancing condition on AL surface; that is, free surface condition is
\begin{equation}
  P_{(In)} n_i   =  \hat{P} n_i + \rho \nu \; (\frac{\partial v_i}{\partial x^k}
  + \frac{\partial v_k}{\partial x^i} ) n_k
- \gamma \frac{d^2 \xi(x)}{d x^2}  n_i.
\end{equation}
The index means $x^1 = x ; \; x^2 = y$  and $n_i$ is the normal unit vector
to the AL surface. $P_{(In)}$ is the pressure just under the AL surface, and
$\hat{P}$ is the atmosphere pressure.
For the above equations, we approximated surface tension term as $\gamma \xi''$ 
by neglecting the second order term in $\mu$.

Now we rewrite the equations in dimensionless form.\\
The incompressible fluid condition is
$$ \frac{\partial u_*}{\partial x_*}+ \frac{\partial v_*}{\partial y_*}=0.$$
To hold this condition automatically, we introduce dimensionless stream function by
\begin{equation}
u_* = \frac{\partial\psi}{\partial y_*}, ~~~v_* = -\frac{\partial\psi}{\partial x_*}.
\end{equation}

In the following, we use the stream function instead of the velocity.
Also, we drop the * mark on the dimensionless quantities.
All the quantities are dimensionless till the end of this section.\\
Now Navier Stokes equation becomes
\begin{eqnarray}
\psi_{yyyy} &=& \mu R [\psi_y \psi_{xyy} - \psi_x \psi_{yyy}]- 2\mu^2 
\psi_{xxyy} \nonumber \\
&& -\mu^3 R[\psi_x \psi_{xxy}-\psi_y \psi_{xxx}] -\mu^4
\psi_{xxxx},
\end{eqnarray}
where the indices indicate derivatives with respect to $x$ and $y$.
The 4th-order derivatives appearing in l.h.s. come from viscous term with taking derivative to cancel out
 the pressure term in equation (\ref{eq:NS}).
The pressure is determined from Navier Stokes equation by using the stream function as follows.
\begin{equation}
P_x = \frac{1}{\mu}+ \frac{1}{2\mu}\psi_{yyy} -\frac{R}{2}(\psi_y \psi_{xy}
- \psi_x \psi_{yy}) + \frac{\mu}{2}\psi_{xxy} ,
\end{equation}
or,
\begin{eqnarray}
P_y &=&  - \cot \theta -\frac{\mu}{2}\psi_{xyy}- 
\frac{\mu^3}{2}\psi_{xxx}\nonumber \\
&&  -\frac{\mu^2 R}{2}(-\psi_{y} \psi_{xx} + \psi_x \psi_{xy}).
\end{eqnarray}

For the stream function we include only 0-th and 1st orders in $\mu$, and for pressure, 
only 0-th order terms are kept.

The Navier-Stokes equation for the stream function and the pressure equations become 
\begin{eqnarray}
\psi_{yyyy} &=& \mu R [\psi_y \psi_{xyy} - \psi_x \psi_{yyy}],\\
P_x &=& \frac{1}{\mu}+ \frac{1}{2\mu}\psi_{yyy}-\frac{R}{2}(\psi_y \psi_{xy}
- \psi_x \psi_{yy}),\\
P_y &=& - \cot \theta.
\end{eqnarray}

Next we consider the boundary conditions up to $\cal{O}(\mu)$.
\begin{eqnarray}
\psi_x(x, y=\phi)&=& \psi_y(x, y=\phi)=0,\\
P(x, y=\xi) &=& \hat{P} -\frac{W_0}{\sin \theta}( \tilde{h}_{xx} - \eta \sin x)
  - \mu \psi_{xy},\\
\psi_{yy}(x, y=\xi)&=& 0,\\
\tilde{h}_x + \eta \cos x &=& - \frac{\psi_x}{\psi_y}(x, y=\xi),
\end{eqnarray}
where we define $W_0 \equiv \mu^2 W \sim 10^{0}$ as order 1,
 because W is $ 7.6 \times 10^2$ in our case.

When $\eta= \tilde{h}=0$, flat laminar flow occurs with the solution
\begin{equation}
\psi = -\frac{1}{3} y^3 + y^2, ~~~~ P = \hat{P} + (1-y) \cot \theta,
\end{equation}
which is easily shown to satisfy Navier Stokes equation and all boundary conditions.

Therefore, we  consider the perturbations from this solution.
The precise perturbative calculations are given in appendix,
and as a result we obtain the height of AL surface,

\begin{equation}
\xi(x) = 1 + \eta \sin x,
\end{equation}
and the stream function given by
\begin{equation}
\psi = -\frac{1}{3}(y-\eta \sin x)^3 + (y-\eta \sin x)^2. \label{eq:stream}
\end{equation}

In our approximation we have  only 0-th order terms in $\mu$.
This is because the first order terms in $\mu$ are also proportional to the small quantity 
$\eta$, the amplitude of a small ripple, which makes them effectively second order quantities.
The form of  stream function is intuitively understood easily.
Because it is just the modification of unperturbative solution for the velocity to be vanished at non flat SL boundary.

\section{Thermal Diffusion Process in the Water Layer}

\subsection{Basic equation}
Now we consider the thermal diffusion process in the fluid that was obtained in the 
previous section. We make the following assumptions.

\begin{enumerate}
\item  We use the long wavelength approximation up to first order in $\mu$
($\mu \equiv kh_0 \sim 6 \times 10^{-2}$).
\item  We ignore thermal expansion of the water and thus retain incompressible fluid.
\item Heat transport is through steady-state thermal diffusion with flow.
\end{enumerate}
 Note that steady-state is valid because the time scale for temperature change is 
much longer  than the time scale for the ice crystal growth. 
The heat flow is given by $ \vec{Q} \equiv -\kappa \vec{\nabla} T + (\rho c
T)\vec{v}$,
where $\kappa$ is the thermal conductivity of water, $T$ the temperature, and  $c$ is the specific heat of water.
The steady-state continuity condition is given by dropping  the time derivative.
  \begin{equation}
  \triangle T - \frac{\vec{v}}{D} \cdot \vec{\nabla} T = 0,\label{eq:thermal}
  \end{equation}
where $D \equiv \frac{\kappa}{\rho c}$ is the thermal diffusivity of water and the incompressibility condition was used.
Furthermore, we dropped the term for the thermal energy coming from energy dissipation of fluid \cite{Landau},
\begin{equation}
-\frac{\rho \nu}{2\kappa}(\frac{\partial v_i}{\partial x_k} + \frac{\partial v_k}{\partial x_i})^2.
\end{equation}
Because this term is much smaller than the other terms.
 In a mean  while we use dimensionless parameters $(x_*, y_*)$ again.
From the previous section,
$$ u= U_0 \frac{\partial \psi}{\partial y_*},~~~~~v = -\mu U_0
\frac{\partial \psi}{\partial x_*},$$
where $\psi$ is the dimensionless stream function.
Then  equation  (\ref{eq:thermal}) becomes
\begin{equation}
 \frac{\partial^2 T}{\partial y_*^2} =
  \alpha[ \frac{\partial \psi}{\partial y_*}\frac{\partial T}{\partial x_*} -
\frac{\partial \psi}{\partial x_*}\frac{\partial T}{\partial y_*}],\label{eq:diff}
\end{equation}
where we have dropped a $\mu^2$ term.
In experimental data, $ D \sim 1.3 \times 10^{-7}$, $ \mu \sim 6 \times 10^{-2}$,
  $U_0 \sim 3 \times 10^{-2}$, $h_0 \sim 10^{-4}$.
  This gives $$ \alpha \equiv \mu \frac{h_0 U_0}{D} \sim 1.4.$$
In the following sections we drop * mark on  the dimensionless quantities again.

\subsection{Expansion in powers of y}
We start from equation (\ref{eq:diff}) with stream function (\ref{eq:stream}).
The temperature at SL boundary will be very nearly equal to the melting temperature
$T_M = 273.15 K$ at atmospheric pressure. 
The surface tension for the curvatures in the experiment (Gibbs-Thomson effect ) 
can alter the melting temperature  by at most $10^{-6}K$, which can be neglected.
Hence we expand the solution in powers of $Y \equiv y-\eta \sin x$.
\begin{equation}
T(x, y) = T_M + a_1(x)Y + a_2(x) Y^2 + \cdots.
\end{equation}
The left and right sides of (\ref{eq:diff}) become
\begin{equation}
T_{yy} = T_{YY}= \sum_{k=2}^{\infty} k(k-1)a_k(x) Y^{k-2},
\end{equation}
\begin{eqnarray}
&& \frac{\partial \psi}{\partial y}\frac{\partial T}{\partial x}
  - \frac{\partial \psi}{\partial x}\frac{\partial T}{\partial y}\nonumber\\
&=& \frac{\partial \psi}{\partial Y}\frac{\partial T}{\partial x}\mid_{Y}
   - \frac{\partial \psi}{\partial x}\mid_{Y}\frac{\partial T}{\partial
Y}\nonumber \\
&=& (2Y - Y^2) \sum_{k=1}^{\infty} \frac{d a_k(x)}{dx} Y^k,
\end{eqnarray}

Setting the left and right sides of (\ref{eq:diff}) equal gives
\begin{equation}
\sum_{k=2}^{\infty} k(k-1)a_k(x)Y^{k-2}= \alpha Y(2-Y)\sum_{k=1}^{\infty}
\frac{d a_k(x)}{dx} Y^k.
\end{equation}
The coefficients are found recursively.
\begin{eqnarray}
a_{n+4} &=& \frac{\alpha}{(n+4)(n+3)} \frac{d}{dx}\{2a_{n+1}-a_n\}, \nonumber \\
&& a_2=a_3=0, ~~a_4 = \frac{\alpha}{6}\frac{d a_1(x)}{dx}.
\end{eqnarray}
All the coefficients are determined when $a_1(x)\equiv a(x)$ is known.
The first nine coefficients are the followings by using  the definition $\hat{D} \equiv \alpha \frac{d}{dx}$.

\begin{eqnarray*}
a_1 &=& a, \\
a_2 &=& 0, \\
a_3 &=& 0, \\
a_4 &=& \frac{\hat{D}}{6} a, \\
a_5 &=& -\frac{\hat{D}}{20} a, \\
a_6 &=& 0, \\
a_7 &=& \frac{\hat{D}^2}{126}a, \\
a_8 &=& -\frac{\hat{D}^2}{210} a, \\
a_9 &=& \frac{\hat{D}^2}{1440}a, \\
&& \\
&\cdots&
\end{eqnarray*}
Because $\alpha \sim O(1)$, we consider only through these second-derivative terms.\\
On the SL surface,
\begin{eqnarray}
T(SL) &=&  T_M = const.\\
Q(SL) &=&  -\kappa \frac{\partial T}{\partial y}\mid_{y= \eta \sin x}= 
-\kappa a(x).
\end{eqnarray}
On the AL surface,
\begin{eqnarray}
T(AL) &=& T_M + [1 + \frac{7}{60} \hat{D} + \frac{13}{3360}
\hat{D}^2] a(x) , \label{ondo}\\
Q(AL) &=& -\kappa[1 + \frac{5}{12} \hat{D} +
\frac{239}{10080} \hat{D}^2] a(x),\label{flow}
\end{eqnarray}
where $Q$ is the heating resulting  from the temperature gradients.
We have omitted the $O(\hat{D}^3)$ terms.
These additional terms are $(6.0 \times 10^{-5}) \hat{D}^3 a$ in (\ref{ondo}),
  and $-\kappa (5.2 \times 10^{-4}) \hat{D}^3 a$ in (\ref{flow}).
To be comparable with $O(\hat{D}^2)$ term, $\alpha$ needs to be about $10^2$.
Therefore, this approximation is valid when $\alpha \ll 10^2$.
To determine $a(x)$, we must consider the temperature and heat flow at the AL surface.

\section{Thermal diffusion in air}
We consider the thermal diffusion in air to consider the temperature and heat flow at the AL surface.
We note two points here.
First, we can not approximate the icicle system as the ramp.
The picture of ramp is a good approximation when we consider inside the thin water-layer, 
but not good for outside. Therefore we treat the icicle as cylindrical object, and consider the thermal diffusion outside.
Second, we can not use the same dimensionless variable as before since our physical space is outside.
Therefore we use different dimensional coordinates in this section.
The diffusion equation in air is given by
\begin{equation}
\Delta T =0.
\end{equation}
Let us write down in cylindrical coordinate.
\begin{equation}
[\frac{\partial^2}{\partial r^2}+ \frac{1}{r}\frac{\partial}{\partial r}+
\frac{1}{r^2}\frac{\partial^2}{\partial \theta^2}+
\frac{\partial^2}{\partial x^2}]T(r,\theta, x) =0.
\end{equation}
We assume axial symmetry, and so we have $\partial T/ \partial \theta =0$.
Therefore, we work with
\begin{equation}
[\frac{\partial^2}{\partial r^2}+ \frac{1}{r}\frac{\partial}{\partial r}+
\frac{\partial^2}{\partial x^2}] T(r, x) =0.
\end{equation}
Because the surface oscillates  in the x-direction, we assume that the solution has the form,
\begin{equation}
T(r,x) = f(r) + g(r)\sin (kx + \phi),
\end{equation}
where we have assumed that the icicle is an infinitely long column with small surface fluctuations.
$f(r)$ satisfies
\begin{equation}
[\frac{d^2}{d r^2}+ \frac{1}{r}\frac{d}{d r}]f(r) =0,
\end{equation}
and $g(r)$ satisfies
\begin{equation}
[\frac{d^2}{d r^2}+ \frac{1}{r}
\frac{d}{d r}-k^2]g(r) =0.
\end{equation}
The solution is,
\begin{equation}
T(r,x) = A + B \log(r/R) + C K_0(kr) \sin (kx + \phi),
\end{equation}
where $K_0$ is the 0-th modified Bessel function, and A, B and C are the constants.
$R$ is the mean radius of the icicle including the thickness of water layer $h_0$. 
Note that the constant C is the order $\delta$, because it is introduced from the fluctuation on icicle.
We define local coordinate $y$ by
$$ r = R + y.$$
The solution is
\begin{eqnarray}
T(x,y) &=& A + B \log(1+y/R) \nonumber \\
&&+ C K_0(k(R+y)) \sin (kx + \phi),
\end{eqnarray}
Note that the ramp system is retained by taking the limit $R \to \infty$ with fixing $B/R$ and $C \exp(-kR)$ finite.

$$ T(x,y) = A + B' y + C' \exp (-ky) \sin (kx + \phi),$$

where B' and C' are other constants.
Hereafter we work with finite $R$ case, because the ramp case is always retained by taking the limit as above.
The reader might think  the appearance of logarithmic term strange, since it diverges at large $r$.
But the appearance of such a term is natural for infinitely long axially symmetric source.
As real icicles have finite length, this solution is valid only close to the icicle; far from the icicle,
 the icicle acts like a point source of heat and 
we must mach the near-icicle and far-icicle solutions at  $r \sim L$ where L is the length of the icicle.
This matching includes the temperature at infinity and partly determines the coefficients A, B and C;
in addition, the mean growth rate of the icicle radius also determine coefficient B, 
which includes the information of the temperature at infinity.

Near the AL surface, $y << R$, we have
\begin{eqnarray}
&&T(x,y) = A + \frac{B}{R}y + \cdots \nonumber \\
&+& C (K_0(kR) + K'_0(kR) ky + \cdots) \sin (kx + \phi)\nonumber \\
&\sim& [A + C K_0(kR)\sin (kx + \phi)] \nonumber \\
&& ~~+ [\frac{B}{R}+C k K'_0(kR) \sin (kx + \phi)] y \nonumber\\
&+& \frac{1}{2}[-\frac{B}{R^2}+C k^2 K''_0(kR) \sin (kx + \phi)] y^2+ \cdots,
\end{eqnarray}
where $K'$ and $K''$ indicate derivative of $K$ with respect  to its argument.
We define the mean growth rate of icicle $V$ by
\begin{equation}
V \equiv -\frac{\kappa_0}{L} < \frac{\partial T}{\partial y}>_{y=0}=
-\frac{\kappa_0 B}{LR},
\end{equation}
where $\kappa_0$ is the thermal conductivity of air, and $< ~~>$ means spatial average over x.
So we obtain
\begin{equation}
B = -\frac{LRV}{\kappa_0}.
\end{equation}
At $y=\delta \sin kx$ (AL surface), the temperature and heat flow are,
\begin{eqnarray}
T(AL) &=& [A + C K_0(kR)\sin (kx + \phi)] \nonumber \\
  && ~ + [-\frac{LV}{\kappa_0}+ C k K'_0(kR) \sin (kx + \phi)] \delta \sin 
kx \nonumber \\
&=& A + [C K_0(kR)\cos \phi -\frac{LV}{\kappa_0}\delta] \sin kx \nonumber \\
&+& C K_0(kR)\sin \phi \cos kx, \label{ondo2}\\
Q(AL)&=& LV - [\kappa_0 C k K'_0(kR) \cos \phi + \frac{LV \delta}{R}] \sin 
kx \nonumber \\
  && ~~~~~~~  - \kappa_0 C k K'_0(kR) \sin \phi \cos kx.\label{flow2}
\end{eqnarray}
Because the constant $C$ is the order $\delta$, we have dropped $\delta C$ terms,
 and we keep terms up to first order in $\delta$.

\section{Growth Rate}
Two boundary conditions apply to the  AL surface: 
continuous temperature and continuous heat flow across the water-air boundary.
Comparing these two sets of equations (\ref{ondo}) and (\ref{ondo2}) , (\ref{flow}) and
(\ref{flow2}), the solution requires that $a(x)= E + F \sin kx + G \cos kx$, where $E,F,$ and $G$ are constants
with dimension of temperature.

Then, in dimensional units, (\ref{ondo}) and (\ref{flow}) are
\begin{eqnarray}
T(AL)&=& T_M + E + [F -\frac{7 \alpha}{60}G - \frac{13 \alpha^2}{3360}F]\sin
kx \nonumber \\
&& ~~~~~~~+ [G +\frac{7 \alpha}{60}F - \frac{13 \alpha^2}{3360}G]\cos kx,
\end{eqnarray}
\begin{eqnarray}
Q(AL)&=& -\frac{\kappa}{h_0} E -\frac{\kappa}{h_0}[F -\frac{5 \alpha}{12}G -
\frac{239 \alpha^2}{10080}F]\sin kx \nonumber \\
&& ~~~-\frac{\kappa}{h_0}[G +\frac{5 \alpha}{12}F - \frac{239 
\alpha^2}{10080}G]\cos kx.
\end{eqnarray}
By comparing to (\ref{ondo2}) and (\ref{flow2}),
we have six equations with six unknowns, ( i.e., A,C,$\phi$,E,F, and G).
A and E can be written immediately:
\begin{equation}
A = T_M - \frac{LV h_0}{\kappa},
\end{equation}
\begin{equation}
E=-\frac{LV h_0}{\kappa}.
\end{equation}
The other four quantities are determined by the following equations.
\begin{eqnarray}
&&\left(
\begin{array}{cc}
1-\frac{13 \alpha^2}{3360} & -\frac{7 \alpha}{60} \\
              \frac{7 \alpha}{60} & 1-\frac{13 \alpha^2}{3360}
\end{array}
\right)
\left(
\begin{array}{c}
F \\
G
\end{array}
\right) \nonumber \\
&& ~~~~~~~~~~=
\left(
\begin{array}{c}
C K_0(kR) \cos\phi - \frac{LV\delta}{\kappa_0} \\
C K_0(kR) \sin\phi
\end{array}
\right),
\end{eqnarray}
\begin{eqnarray}
&&\left(
\begin{array}{cc}
1-\frac{239 \alpha^2}{10080} & -\frac{5 \alpha}{12} \\
                  \frac{5 \alpha}{12} & 1-\frac{239 \alpha^2}{10080}
\end{array}
\right)
\left(
\begin{array}{c}
F \\
G
\end{array}
\right) \nonumber \\
&&~~~~~~~~~~=
\left(
\begin{array}{c}
\frac{\kappa_0}{\kappa} \mu K'_0(kR) C \cos\phi + \frac{LVh_0\delta}{R
\kappa} \\
\frac{\kappa_0}{\kappa} \mu K'_0(kR) C \sin\phi
\end{array}
\right).
\end{eqnarray}
Solving above equations, we obtain
\begin{eqnarray}
F &=& \mu \frac{LV\delta K'_0}{\kappa K_0}\frac{1-\frac{239}{10080}\alpha^2}
{1+ \frac{1272}{10080}\alpha^2 +(\frac{239}{10080})^2 \alpha^4} ,\\
G &=& - \mu \frac{LV\delta K'_0}{\kappa K_0}\frac{\frac{5}{12}\alpha}
{1+ \frac{1272}{10080}\alpha^2 +(\frac{239}{10080})^2 \alpha^4} ,
\end{eqnarray}
where we have neglected second order terms in $\mu$ and we neglected a term proportional to 
$h_0/R (\sim10^{-3})$.

At the SL surface, the growth rate of ice is given by
\begin{equation}
v(x) = \frac{Q(SL)}{L}= V -\frac{\kappa}{Lh_0}(F \sin kx + G \cos kx).
\end{equation}

The form of growth rate is different to usual MS theory.
In MS theory, $v(x) = V + f \sin kx$ for the surface $y = \delta \sin kx$.
But now we have another term $\cos kx$.
To understand its physical meaning,
we write the relative growth rate $v_s$ as the growth rate  in a reference frame moving with velocity $V$
($v_s \equiv v - V$).
\begin{eqnarray}
&& v_s(x) = f \sin kx- g \cos kx,\\
&& f \equiv -\frac{\kappa}{Lh_0}F, ~~~~g \equiv \frac{\kappa}{Lh_0}G.
\end{eqnarray}
The steady-state condition means,
$$v_s(x) \equiv \frac{d y_s(x,t)}{dt}\mid_{t=0},$$ where $y_s$ is the hight of SL surface in reference frame.
From the steady-state condition, the time scale for growth of fluctuation is very long compared to our observing time scale.
By solving the equation:
\begin{eqnarray}
v_s(x) &\equiv& \frac{d y_s(x,t)}{dt}\mid_{t=0}= f \sin kx- g \cos kx,\\
with  && y_s(x,t=0) = \delta \sin kx,
\end{eqnarray}
we obtain 
\begin{equation}
y_s(x,t) = \delta(t) \sin (kx-\omega t),
\end{equation}
with relations
\begin{equation}
\dot{\delta} = f, ~~~~~\omega = g/\delta.
\end{equation}
The essential point is that the fluctuation is not only growing up, but also traveling downwards ($f>0, g>0$).

Therefore the amplification factor is determined from F.
By using the relation $-K'_0 = K_1 > 0$, we have
\begin{eqnarray}
\frac{\dot{\delta}}{\delta}&=& -\frac{\kappa}{Lh_0 \delta} F = Vk
\frac{\frac{K_1(kR)}{K_0(kR)}(1-\frac{239}{10080}\alpha^2)}
  {1+ \frac{1272}{10080}\alpha^2 +(\frac{239}{10080})^2 \alpha^4}.
\end{eqnarray}
Note that $\frac{K_1(kR)}{K_0(kR)}\sim 1+\frac{1}{2kR}$.
  In the case of $kR >> 1/2$, meaning a thick icicle, we obtain
\begin{equation}
  \frac{\dot{\delta}}{\delta} = Vk\frac{1-\frac{239}{10080}\alpha^2}{1+
\frac{1272}{10080}\alpha^2 +(\frac{239}{10080})^2 \alpha^4}.\label{AF}
\end{equation}

Because $\alpha \propto k$, this form is similar to the amplification factor
  given by Mullins-Sekerka theory.
The  amplification factor increases proportional to wave number by thermal diffusion in air as expected for a Laplace 
instability, and it decays by interaction with fluid $(\alpha terms)$.
The thermal diffusion in thin water flow works just like Gibbs-Thomson effect,
since the fluid changes temperature distribution to be uniform and inhibit Laplace instability.
 From these two effects, we have maximum value for $\dot{\delta}/\delta$.

\begin{figure}[t]
\begin{center}
\includegraphics[width=5cm]{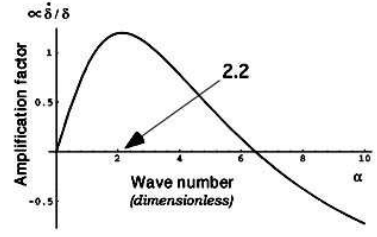}
\end{center}
\caption{The dimensionless amplification factor $\frac{h_0^2 U\dot{\delta}}{DV\delta}$
versus the dimensionless wave number $\alpha$. Laplace instability in air
and a GT like effect due to hydrodynamics amplifies perturbations
 that have a wavelength given by an $\alpha$ close to 2.2.}
\end{figure}

The maximum amplification factor occurs when $\alpha = 2.2$, which determines a preferred wavelength.
  $$\alpha_{max} \sim 2.2.$$
By using  $D = 1.3 \times 10^{-7}m^2/s$  with experimental data:
$$U_0 = (2.4 \sim 4) \times 10^{-2} m/s,~~~h_0 = (0.93 \sim 1.21) \times 10^{-4} m,$$
we have
\begin{equation}
\lambda_{\max} = \frac{2\pi}{k_{max}}= 2\pi \frac{h_0^2 U_0}{D 
\alpha_{max}}= 5mm \sim 13mm,
\end{equation}
which well agrees with experimental value of 8 mm. \cite{GR}
Also, in agreement with observations, this value does not directly depend on external 
temperature.

Using the constraints between $U_0$, $h_0$, and flow quantity $Q$ as,
\begin{equation}
U_0 = \frac{gh_0^2 \sin \theta}{2 \nu},\; \; \; Q = \frac{4\pi}{3} R U_0 h_0.
\end{equation}

Then $Q$ is related to $\lambda_{\max}$ as
\[ \lambda_{\max} = \frac{1}{D \alpha_{max}}(\frac{\nu}{g 
\pi})^{1/3}(\frac{3Q}{2R})^{4/3},\]
where $\theta = \pi/2$ is assumed.

Therefore if $Q$ is proportional to $R$ for usual icicles, its wave length is uniquely determined.
But for thin icicles with $R \le \lambda/(4\pi)$,
we should include the  $\frac{1}{2kR}$ term that we have neglected.

Next we consider the travel of fluctuations along the icicle.
The traveling phase velocity $w$ is the following function of $\alpha$.
\begin{equation}
w \equiv \frac{\omega}{k} = \frac{\kappa G}{L\delta h_0 k}
= \frac{K_1 V}{K_0} \beta(\alpha) \sim V \beta(\alpha),
\end{equation}
where $\beta(\alpha)$ is defined by
\begin{equation}
\beta(\alpha) \equiv  \frac{\frac{5}{12}\alpha}
{1+ \frac{1272}{10080}\alpha^2 +(\frac{239}{10080})^2 \alpha^4},
\end{equation}
and its form is given in Figure 5.

\begin{figure}[t]
\begin{center}
\includegraphics[width=5cm]{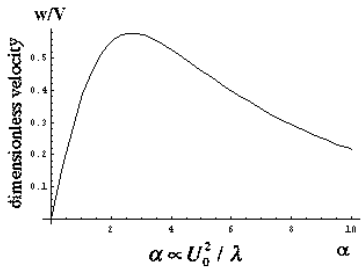}
\end{center}
\caption{The dimensionless speed $w/V$ that the ripples travel down an icicle 
versus $\alpha \sim U_0^2/\lambda$, where V is the growth rate of icicle.}
\end{figure}

At the 8-mm wavelength, fluctuation travels downwards very slowly with 
speed $w \sim 0.5 V$, where $V$ is the mean growth rate ($V = \dot{R}$) .
For the ambient air temperature of -8 degrees centigrade, this speed is about 1 mm per hour in experiment .
Because water flow carries heat flow downwards, this kind of motion seems to be natural.
Actually the speed $w=0$ when $U_0 =0$ ($\alpha =0$).
Then we can imagine that such speed will be increasing function of fluid velocity $U_0$.
But this is not true.
For (solid) surface fluctuation to change its form, it is necessary to emit 
latent heat non uniformly.
Like the amplification (growing) factor becomes maximum at the specific 
value of $\alpha$,  the traveling speed may have such  dependence of $\alpha$.
Qualitatively this is due to the fact that fluid not only to carry heat downwards (small $\alpha$ region in fig.5),
  but also make the temperature field uniform and thus suppresses the heat diffusion (large $\alpha$ region in fig.5).

\section{Conclusions}

We have shown that an icicle covered with thin water flow makes wave-like undulations on the ice 
during solidification, and the preferred wavelength is determined by a MS-like theory.
Thermal diffusion in air makes wavelength shorter, i.e. the amplification 
factor becomes larger for shorter wavelengths.
On the other hand, thermal diffusion in thin water flow makes the wavelength larger, i.e. 
the amplification factor becomes smaller for shorter wavelengths.
From these two effects,  a specific wavelength emerges with a maximum amplification factor of fluctuation.
The thermal diffusion in the thin water flow works just like the 
Gibbs-Thomson effect because the water flow makes the temperature distribution more
uniform and thus inhibits the Laplace instability.  This then is one of our main results. 
\

Our $ \lambda_{\max}$ depends on $\theta$ and flow quantity $Q \equiv 2 \pi 
R  h_0 \bar{U} = l h_0 \bar{U} $.
 ($\theta$ dependence will be observed for flow on inclined gutter experiment: Fig2.
For an icicle, $\theta = \pi/2$ should be used.)

\begin{equation}
\lambda_{\max} = \frac{1}{D \alpha_{max}}(\frac{\nu}{g \pi \sin 
\theta})^{1/3}(\frac{3Q}{2R})^{4/3},
\end{equation}

where we have used the relative Nusselt equation \cite{Landau},
\begin{equation}
  h_0 = [\frac{\nu Q}{g \pi R  \sin \theta}]^{1/3}.
\end{equation}

The $\theta$ dependence of the wavelength for the ramp case is discussed by Matsuda \cite{matsu} experimentally as
\begin{equation}
\lambda \sim \frac{8.2}{\sin^{\gamma}\theta}[mm],
\end{equation}
with $\gamma = 0.6 \sim 1$.
However, the number of plotting data is few, and thus we can not give a definite result experimentally.
$\gamma$ is 0.3 in our theory. The difference exists, but not a big disagreement qualitatively.

In nature, all icicles have their own flow rates $Q$,
  but almost all have the ripples with the same wavelength.
This is explained by our analysis because  the wavelength depends on ratio of $Q$ to $R$, but not only on $Q$ itself.
It is then natural to assume that in nature, $Q$ is proportional to $R$ .
The selected wavelength of our analysis does not depend explicitly on external temperature.
Although the mean growth rate $V$ and also the amplification factor increase with a decrease of temperature, 
 the selected wavelength with the maximum amplification factor is independent of icicle growth rate (\ref{AF}).

We have also shown that surface ripples are expected to travel downwards during icicle growth 
with a  speed of 0.5 times the average normal growth rate of the ice.
This should be checked by experiment.
Our theory may be used to map the waves around mineral stalagmite by changing
the diffusion equation from temperature field to a solute density field.
Furthermore, our diffusion equation in fluid is mathematically similar to
the Schr\"odinger equation for a harmonic oscillator having the complex valued 
potential. Therefore, the algebraic method may be possible to be used for analysis.
This point will be discussed elsewhere.

\begin{acknowledgments}
The authors are grateful to Prof. R. Takaki, and Dr. P.L. Olivier for the valuable discussion.
One of the authors N.Ogawa thanks Prof. K. Fujii for the continual encouragement.
The authors would like to thank Prof. E. Yokoyama, Dr. Nishimura, and Prof. R. Kobayashi
for the helpful discussions and encouragement, and to Ms K. Norisue for her help in English corrections.
\end{acknowledgments}

\section{Appendix}
We define the fluctuation fields $\tilde{\psi}$ and $\tilde{P}$ by the 
following relation:
\begin{eqnarray}
\psi &=& -\frac{1}{3} y^3 + y^2 + \tilde{\psi}(x,y),\\
P &=& (1-y) \cot \theta + \hat{P} + \tilde{P}.
\end{eqnarray}

Then for the fluctuation fields, we have following equations:
\begin{eqnarray}
\tilde{\psi}_{yyyy} &=& \mu 
R[(-y^2+2y+\tilde{\psi}_y)\tilde{\psi}_{xyy}\nonumber \\
  && -(-2+ \tilde{\psi}_{yyy})\tilde{\psi}_x],\label{NS}\\
\tilde{P}_x &=& \frac{1}{2\mu}\tilde{\psi}_{yyy}
-\frac{R}{2}[(-y^2+2y+\tilde{\psi}_y)\tilde{\psi}_{xy} \nonumber \\
&& -(-2y+2+\tilde{\psi}_{yy}) \tilde{\psi}_x],\label{Px}\\
\tilde{P}_y &=& 0.\label{Py}
\end{eqnarray}

The boundary conditions are
\begin{eqnarray}
\tilde{\psi}_x (SL)&=&0,\\
\tilde{\psi}_y (SL)&=& \eta^2 \sin^2 x -2\eta \sin x,\\
\tilde{P}(AL) &=& -\frac{W_0}{\sin \theta}(\tilde{h}_{xx}- \eta \sin 
x)\nonumber \\
&& +(\tilde{h} + \eta \sin x) \cot \theta,\label{BC.AL1}\\
\tilde{\psi}_{yy}(AL) &=& 2(\tilde{h}+\eta \sin x),\label{BC.AL2}\\
\tilde{\psi}_{x}(AL) &=& (\tilde{h}_x +\eta \cos x) \times [(1+\eta \sin x 
+ \tilde{h})^2 \nonumber \\
&& - 2(1+\eta \sin x + \tilde{h})-\tilde{\psi}_y(AL)],\label{BC.surf}
\end{eqnarray}
where the AL surface is determined by $y=1+\eta\sin x +\tilde{h}$,
and the SL surface is determined by $y= \eta \sin x$.

To solve the above equations, we assume the solutions have the form,
\begin{eqnarray}
&\tilde{\psi}& = \tilde{\psi}^{(0)} + \mu \tilde{\psi}^{(1)} + \cdots,\\
&\tilde{P}& = \tilde{P}^{(0)} + \cdots,\\
&\tilde{\psi}^{(0)}_{yyy}& = 0,
\end{eqnarray}

and neglect higher order in $\mu$.
The solution for the 0-th ordered stream function and pressure are
\begin{eqnarray}
\tilde{\psi}^{(0)} &=& (\tilde{h}+\eta \sin x)y^2 - [\eta^2 \sin^2 x 
+2(1+\tilde{h})\eta \sin x] y \nonumber \\
&&+ [(1+\tilde{h})\eta^2 \sin^2 x + \frac{1}{3} \eta^3 \sin^3 x],\\
\tilde{P}^{(0)}&=& -\frac{W_0}{\sin \theta}(\tilde{h}_{xx}-\eta \sin
x)+(\tilde{h}+\eta \sin x)\cot \theta.\label{P0}
\end{eqnarray}

To obtain the 1st-order stream function, we put above solution into the 
following equation.
\begin{eqnarray}
\tilde{\psi}^{(1)}_{yyy} = &-&\frac{2W_0}{\sin \theta} (\tilde{h}_{xxx} -
\eta \cos x)+2(\tilde{h}_x + \eta \cos x)\cot \theta  \nonumber \\
+R[(-y^2&+&2y+\tilde{\psi}^{(0)}_y)\tilde{\psi}^{(0)}_{xy} - (-2y + 2
+ \tilde{\psi}^{(0)}_{yy})\tilde{\psi}^{(0)}_x],
\end{eqnarray}

which is obtained by (\ref{Px}) with the help of (\ref{P0}).
This expression is consistent with (\ref{NS}).
After some integrations with the boundary conditions, we obtain
\begin{eqnarray}
\tilde{\psi}^{(1)}&=&
\frac{R}{30}(1+ \tilde{h})\tilde{h}_x ~y^5
-\frac{R}{6}\eta(1+ \tilde{h})\tilde{h}_x \sin x ~y^4 \nonumber \\
&+&\frac{R}{3}\eta^2(1+ \tilde{h})\tilde{h}_x \sin^2 x ~y^3 \nonumber \\
&+&\frac{1}{3}[-\frac{W_0}{\sin \theta}(\tilde{h}_{xxx}-\eta \cos x)
+(\tilde{h}_{x}+\eta \cos x)\nonumber \\
&\times& \cot \theta] ~y^3 + \frac{1}{2}C_1(x) y^2 + C_2(x) y + C_3(x).
\end{eqnarray}
$C_1,C_2,C_3$ are determined by the boundary conditions.
The result are

\begin{eqnarray}
C_1 &=& -2(1+\tilde{h}+\eta \sin x)[R(1+\tilde{h})\tilde{h}_x \nonumber \\
&\times & \{\frac{1}{3}(1+\tilde{h}+\eta \sin x)^2- (1+ \tilde{h}) \eta 
\sin x \}\nonumber \\
&-& \frac{W_0}{\sin \theta}(\tilde{h}_{xxx}-\eta \cos x) \nonumber\\
&+&(\tilde{h}_x + \eta \cos x)\cot \theta],\\
C_2 &=& -\frac{1}{2}R(1+\tilde{h})\tilde{h}_x \eta^4 \sin^4 x \nonumber\\
&+& [\frac{W_0}{\sin \theta}(\tilde{h}_{xxx}-\eta \cos x)-(\tilde{h}_x + 
\eta \cos x) \nonumber \\
&\times&\cot \theta]\eta^2 \sin^2 x  - C_1(x)\eta \sin x,\\
\frac{\partial C_3}{\partial x}&=& -\frac{R}{5}[(1+ 
\tilde{h})\tilde{h}_x]_x \eta^5 \sin^5 x \nonumber\\
&-&\frac{R}{2}(1+ \tilde{h})\tilde{h}_x \eta^5 \sin^4 x~\cos x \nonumber \\
&+& \frac{\eta^3}{3}[\frac{W_0}{\sin \theta}(\tilde{h}_{xxxx}+\eta \sin x) 
\nonumber \\
&-&(\tilde{h}_{xx}- \eta \sin x)\cot \theta] \sin^3 x \nonumber \\
&-& \frac{\eta^2}{2}\partial_x C_1(x) \cdot \sin^2 x - \eta \partial_x 
C_2(x) \cdot \sin x.
\end{eqnarray}

Now we put $\tilde{\psi} = \tilde{\psi}^{(0)} + \mu \tilde{\psi}^{(1)}$ into
(\ref{BC.surf}). We now have,
\begin{eqnarray}
&&2(1+\tilde{h})^2 \tilde{h}_x + \mu[\tilde{\psi}^{(1)}_x(AL) \nonumber \\
&+& (\tilde{h}_x + \eta \cos x) \tilde{\psi}^{(1)}_y(AL)]=0.
\end{eqnarray}

We expand the fluctuation of water-layer thickness as
\begin{equation}
\tilde{h} = \tilde{h}^{(0)}+ \mu \tilde{h}^{(1)},
\end{equation}
and by putting it into above equation, we have $\tilde{h}^{(0)}_x=0$, and 
so we can use
\begin{equation}
\tilde{h}^{(0)} =0.
\end{equation}

This can be done by redefining of $h_0$ after subtracting a constant.
Then $\tilde{h}^{(1)}$ is determined by
\begin{equation}
\tilde{h}^{(1)}_x = -\frac{1}{2}[\tilde{\psi}^{(1)}_x(AL) + \eta \cos x
\tilde{\psi}^{(1)}_y(AL)]_{\tilde{h}=0}.\label{hight}
\end{equation}

 From the fact that $\tilde{h}$ starts from ${\cal O(\mu)}$ in its expansion,
we can determine $\tilde{\psi}^{(1)}$  as ,
\begin{eqnarray}
\tilde{\psi}^{(1)}&=& (\frac{W_0}{\sin \theta}+\cot \theta)
[\frac{1}{3}\eta \cos x \cdot y^3\nonumber \\
&-& \eta \cos x (1+\eta \sin x)y^2 +\{\eta^3 \sin^2x \cos x \nonumber \\
&& ~~~~~+ 2\eta^2 \sin x \cdot \cos x \}y] + C_3,\\
\partial_x C_3 &=& \eta^3 \sin x (\frac{W_0}{\sin \theta}+\cot 
\theta)\nonumber \\
&\times &[\frac{4}{3} \eta \sin^3 x - \eta \sin x + 3\sin^2x -2].
\end{eqnarray}

 From the above expression and (\ref{hight}), we obtain the following simple relation:
\begin{equation}
\tilde{h}^{(1)}_x = -\frac{\eta}{3}(\frac{W_0}{\sin \theta}+\cot \theta)\sin x.
\end{equation}

Furthermore,
\begin{equation}
\tilde{h}^{(1)} = \frac{\eta}{3}(\frac{W_0}{\sin \theta}+\cot \theta)\cos x.
\end{equation}

The height of the AL surface is,
\begin{equation}
\xi(x) = 1 + \eta \sin x ,
\end{equation}
whereas the stream function is
\begin{eqnarray}
\psi &=& -\frac{1}{3}y^3 + y^2 + \tilde{\psi}^{(0)}(\tilde{h}) + \mu
\tilde{\psi}^{(1)},\nonumber \\
&=& -\frac{1}{3}(y-\eta \sin x)^3 + (y-\eta \sin x)^2.
\end{eqnarray}


\begin{thebibliography}{99}
\bibitem{GR}
The amplification rate of a wave (fluctuation)  depends on its wavelength.
Only the wave with the largest amplification rate can survive during the 
development of the fluctuations,  other modes will be observed as lower-amplitude noise.
The constant rate of $\dot{\delta}/\delta$ means that the growth is exponential in time; hence,
small differences of this constant results in large differences in fluctuation amplitude at long times.
In this sense,  a mode distribution like that in Fig. 4 will be hidden in the external noise
and thus not observed in experiments. This situation is the same as in the original MS theory.
\end{thebibliography}
\end{document}